\documentstyle[12pt,aasms]{article}

\def\ga{\mathrel{\mathpalette\fun >}}

\def\fun#1#2{\lower0.837ex\vbox{\baselineskip0ex\lineskip0.209ex
  \ialign{$\mathsurround=0ex#1\hfil##\hfil$\crcr#2\crcr\sim\crcr}}}

\def\msun{M_\odot}

\def\sles{\lower2pt\hbox{$\buildrel {\scriptstyle <}
   \over {\scriptstyle\sim}$}}
 
\def\sgreat{\lower2pt\hbox{$\buildrel {\scriptstyle >}
   \over {\scriptstyle\sim}$}}

\def\ga{\mathrel{\mathpalette\fun >}}

\begin{document}

\title{ On the Role  of Irradiation   and Evaporation in Strongly
   Irradiated Accretion Disks in the Black Hole X-ray Binaries:
   Toward an Understanding of FREDs and Secondary Maxima }

\smallskip

\author{  John K. Cannizzo\footnote
{ Universities Space Research Association }}
%
\affil{e-mail: cannizzo@stars.gsfc.nasa.gov}
\affil
{NASA/GSFC/Laboratory for High Energy Astrophysics, 
  Code 662, Greenbelt, MD 20771}
\authoraddr
{NASA/GSFC/Laboratory for High Energy Astrophysics, 
Code 662, Greenbelt, MD 20771}

\smallskip

 \vskip 2truein

\centerline{ to appear in the Astrophysical Journal Letters,
  2000, May 1, vol. 534}
\received{ 2000 February 22 (Revised: March 14)  }
\accepted{ 2000 March 20 }

\begin{abstract}

We examine   a new paradigm to
account for the exponential
decay seen in the light curves
of some of the 
              bright X-ray novae.
    These systems
show an exponential decay in soft X-rays
with an $e-$folding time constant of $\sim30$ d.
   We investigate a scenario in which
evaporation of  matter into a corona
is the dominant mass removal mechanism
from the accretion disk.
  We utilize the  thermal evaporative
instability discovered by Shaviv and Wehrse.
   First we  parametrize  local mass loss
rates from the disk 
 (fitted to vertical structure
computations of the optically thin structure
using the photoionization code
    CLOUDY), and then we utilize the 
scalings  in our numerical
   time dependent model
  for the decay. Both the $\sim30$ d $e-$folding 
time scale for the decay and the secondary maximum
with its rapid rise time $\sim1-3$ d
which is seen in the
   X-ray nova  light curves can be produced
by adjusting the strength of the evaporation.

\end{abstract}

\medskip

{\it  Subject headings:}
accretion,  accretion disks $-$ 
           black hole physics  $-$ transients: individual (A0620-00)

\section{ INTRODUCTION }


The accretion disk  limit cycle
model has been successful
in explaining the outbursts
seen in dwarf novae
and X-ray transients
(see Cannizzo 1993a, 1998,
     and references therein;
 Hameury et al. 1998).
    In several of the brightest and
best studied X-ray novae one sees a rapid rise
to maximum light, followed by an exponential 
decline. The time constant associated with the decline is
$\sim30-40$ d   (Mineshige et al. 1993, Chen et al. 1997).
  In addition, one also sees a secondary maximum $\sim50$ d
after the initial
     maximum, during which time the X-ray flux increases
by a factor $\sim1.5-2$ before resuming its exponential decay.
               The post-secondary maximum light curve
  is therefore  parallel to (but offset from) the pre-secondary
   maximum light curve when plotted as $\log L_X$ versus  time.

  The X-ray novae are thought to consist
of interacting binaries in which a mass losing
late-type star transfers matter onto a black hole (BH)
(Tanaka \& Shibazaki 1996).
If we adopt the canonical wisdom that the
X-ray flux seen in the outbursts is due to accretion
from a transient accretion disk, then logic would dictate
that the reason for the decline in X-ray flux is a
decline in the mass of the accretion disk.
The question then arises as to where the dominant
mass removal occurs.
Three natural  possibilities present themselves
for removing gas from the viscously
active inner disk:
  (1) outflow at the outer edge due to 
thermal instability, (2) standard viscous
 accretion onto the BH, and (3) evaporation
of gas from the surface of the disk.

{\it 1.  mass removal from the outer disk edge:}
Cannizzo et al.  (1995) proposed that the exponential decays
are due to the action of a cooling front in the disk.
   A cooling front in the limit cycle model (Cannizzo 1993a)
invariably begins at large radii in the accretion disk
and moves to smaller radii, producing a vigorous outflow
at the hot/cold interface  that is several times as large
as the  rate of accretion onto the central object from the
  inner disk (Vishniac \& Wheeler 1996, Vishniac 1997;
 Menou, Hameury, \& Stehle 1999).
For this scenario to be workable in the context of the X-ray
novae, however,
       the   viscosity parameter for ionized gas
        $\alpha_{\rm hot}$ must have
the form $\alpha_{\rm hot}\simeq \alpha_0(h/r)^n$, where 
$\alpha_0\simeq 50$ and $n\simeq 1.5$. There are several problems with
this.  First,  
      in the outbursting dwarf novae we infer $\alpha_{\rm hot}\simeq0.1$
from the Bailey relation between decay time and orbital period (Smak 1984,
   1999),
therefore one would have to invoke a special form for $\alpha_{\rm hot}$
in the X-ray novae.
  (The value $\alpha_{\rm hot}\simeq0.1$
exceeds by a nominal factor
      estimates 
   taken from  three-dimensional  MHD calculations of the nonlinear
       saturation limit for the
Balbus-Hawley instability,
  the currently favored model for
  generating viscous dissipation from the Keplerian
  shear in accretion disks
                  [Stone et al. 1996, Fleming et al. 2000]).
 Second,  the strong irradiation in outbursting
X-ray novae should prevent the cooling front from forming. 
If as little as $\sim10^{-3} - 10^{-4}$ of the central X-ray flux is 
   received by 
the accretion disk via
   indirect  scattering, the locally defined irradiation
temperature would exceed $\sim10^4$ K at the outer disk edge, 
which would be sufficient to keep the whole disk ionized.
%

{\it 2. mass removal from the inner disk edge:}
           It has been known for some time that irradiation
is a much stronger physical effect in the LMXBs
than in the cataclysmic variables 
         (van Paradijs \& Verbunt 1984, van Paradijs 1996).
  As just noted, 
    strong irradiation of the outer portions of the 
accretion disk by X-ray flux coming from the inner disk
edge will tend to keep the disk in a hot, ionized state,
and prevent a cooling front from forming (Tuchman et al. 1990,
  Mineshige et al. 1990).  In this case the decay
rate will be slower,   for a given $\alpha$ law,
insofar as viscous processes operate on a longer time scale
than thermal processes.
     King
     \& Ritter (1998, see also King 1998, Shahbaz et al. 1998) 
     argue for   a viscous decline
to account for the $\sim30$  d $e-$folding decays in the
X-ray novae,  because
one does not expect a cooling  front to be present.
This scenario is not without difficulties, however. If we assume that 
 $\alpha_{\rm hot}\simeq0.1$ as for  outbursting dwarf nova
disks,
then for an accretion disk with an outer
  radius $r_{\rm outer} \approx 10^{11}$ cm relevant for
    A0620-00, for example, the locally
defined $e-$folding decay time in the X-ray light curve would be
     $\sim300$  d   $-$ about a factor of ten slower than observed.
  In addition, Kaluzienski et al. (1977)
report a slight  increase in the rate of decay of the 1975
outburst of A0620-00 during the later stages of
the outburst.
 They fit an $e-$folding decay time scale of
$\sim29$ d to the   pre-secondary maximum light curve,
and a decay time of $\sim 21$ d to the post-secondary 
maximum light curve.
 A viscous decay would produce the opposite
effect $-$ an increase in the viscous time scale due
to decreased mass in the disk as the outburst proceeds.
  This comes about because the viscous
time scale
    $\tau_v = [\alpha\Omega (c_s/r\Omega)^2]^{-1}$
  varies as the reciprocal of the midplane temperature
$T_{\rm midplane}$ ($\propto c_s^2$),
and as the disk drains onto the BH, the surface
densities  and concomitant midplane temperatures diminish.


{\it 3.  mass removal from  intermediate radii:}
Shaviv \& Wehrse (1986, hereafter SW86) discovered an evaporative instability
associated with the local minimum in the pressure in the 
optically thin 
vertical structure of an  accretion disk overlying 
                  the photosphere.
  Recently de Kool \& Wickramasinghe (1999, hereafter dKW) computed
the amplitude of this effect for both non-irradiated systems
(dwarf novae) and strongly irradiated systems (X-ray novae).
They found that evaporation can be dominant for the latter systems.
dKW present a simple formalism for using a photoionization
code to calculate the vertical structure.
 We note that other physical mechanisms have been discussed
for removing disk material in the form of a wind $-$
for instance the ``coronal siphon flow'' (Liu,  Meyer, \& 
Meyer-Hofmeister 1995, 1997). In this work 
  we  utilize the SW86 mechanism because
   it is conceptually straightforward and  one can readily
quantify its strength in simple vertical structure calculations.
 Many previous workers have
  considered strongly irradiated
  disks 
(e.g. Begelman et al. 1983;
    Ko \& Kallman  1991;  Raymond 1993;
 Hameury et al. 1997, Dubus et al. 1999, Menou et al. 2000).

We perform vertical structure calculations using the
photoionization
    code CLOUDY
 (Ferland 1996) for an accretion disk
around  a  $10\msun$ black hole, 
    and parametrize conditions at the local minima
in pressure $-$ the point at which hydrostatic equilibrium
can no longer   be maintained. The local outflow rate
in a given annulus is expressed in terms of power law scalings
in $r$ and ${\dot M}$,
and these are then utilized in our global   time
dependent computations to calculate light curves
 based on the rate of accretion onto the central object,
and the $V$ band flux from the accretion disk.

\section {  MODEL CALCULATIONS }

As in dKW, we simply set the heating and cooling
rates to be equal in the photoionization code.
   The heating has two sources:
      X-ray irradiation  from accretion
      onto the central object, and
      local viscous 
       dissipation. The first term is by far the dominant
one in this work.
   Cooling  occurs through a variety  of physical
    processes
 calculated by CLOUDY.
 Our method is the same as  that described
in Section 4.2 of dKW (except that we use
    CLOUDY rather than MAPPINGS).
 The heating rate per unit volume
is $1.5 \alpha \Omega P_g$,
where $\alpha_{\rm chromosphere}   = 0.3$, $\Omega$ is the
local Keplerian angular velocity,
and $P_g$ is the gas pressure.  
   The irradiation heating comes via
a dilute radiation field,
$T_{bb} = T_{\rm d, eff}(r_{\rm max})$,
where $T_{\rm d, eff}$ is the effective temperature
given by $\sigma T_{\rm d, eff}^4 = (3/(8\pi)) \Omega^2 {\dot M}$
(Shakura \& Sunyaev 1973),
and $r_{\rm max}$ is the radius where most of the
energy is dissipated.
 The radiation field is diluted by a geometric factor
$\eta r_{\rm max}^2 / (4  r^2)$, where  $\eta  = 0.1$,
   and $r_{\rm max}=10^7$ cm.
Thus we assume the irradiation to be indirect
rather than direct.
  For 
      BH parameters, the photosphere
of the accretion disk is irradiation-dominated
at all radii (see eqn. [3] of Ko \& Kallman 1991).

 Figure  1 shows the 
vertical structure of the optically thin part of
the accretion disk overlying the photosphere
of an optically thick disk 
  for five values of  ${\dot M}$,
      at one representative
radius  in the disk.
   For the three lowest   ${\dot M}$ 
  curves
    in the first panel
      one can see 
a local minimum in the pressure.
  The third panel  shows 
   $2\pi r^2 \rho c_s^2$. The value of this
quantity evaluated  at the minimum in pressure estimates
the local outflow (Czerny \& King 1989).
   We utilize power law scalings of these minima
(as functions of  $r$   and ${\dot M}$)
in our time dependent computations to calculate
the local mass loss rates. 
 In our  time dependent 
     model, we  take the local mass loss
rate from a given annulus  $\Delta{\dot M}(r)$
 to be $4\pi r \Delta r \rho c_s^2$.
%
%
%
%
%
%
%
%
   Our
    scaling based on the CLOUDY results
is  $\Delta{\dot M}(r) = 2 A \Delta r /r$,  where
$A$ is the local wind loss rate parametrization
as considered by dKW and shown in Fig. 1, 
namely 
$2\pi r^2 \rho c_s^2$ which we approximate as $10^{17}$ g s$^{-1}$
${\dot M}_{c, 18}^{1/4}$.
The quantity ${\dot M}_{c, 18}$ is the rate of mass loss from
the inner accretion disk onto the BH in units of 
    $10^{18}$ g s$^{-1}$.
%
  One  caveat to note:
   CLOUDY is not intended to be accurate
for densities significantly above about $10^{-10}$ g cm$^{-3}$.
Although the densities corresponding to the minima in $P_{\rm gas}$
for the specific radius $10^{9.5}$ cm
     shown in Fig. 1 are less than this,
disk densities generally increase with decreasing radius,
so that for smaller radii our scaling for
     $\Delta{\dot M}(r) $  may begin to break down.
In this inner disk region other deficiencies 
in our time dependent
    model  connected with our neglect  of
  general   relativistic effects
  such as transonic flow also become manifest.
We plan to account for these effects more fully in future work. 
Another caveat is our assumption that a constant fraction
of the indirect irradiation is received by the disk as
the outburst fades, which we make for simplicity.
  A multidimensional treatment would be required
      in order to make a   more self-consistent model.
 Finally, we cannot address the ultimate (nonlinear)
     fate of the gas lost
in the wind. In this work we simply assume that it becomes added
to a static corona
 which is dynamically 
    decoupled from the underlying accretion disk.

The time dependent code we utilize to follow the accretion
disk evolution is a modification of one described extensively
in earlier works (Cannizzo 1993b, Cannizzo et al. 1995,
Cannizzo 1998). The two modifications in this work
are that (1) we allow for the effects of indirect
irradiation, using the form cited earlier. The effect
of irradiation on the   steady state solutions
was taken from Tuchman et al. (1990).
(2) We also remove matter at each annulus in accord
with the aforementioned 
$\Delta{\dot M}(r)  $ prescription.
    All aspects of the code which involve
   thermal transition fronts
  are not utilized in this work, because irradiation
    prevents the cooling front from forming
and keeps the entire disk in the ionized state.
We assume a central BH mass of $10\msun$,
  and we distribute 1000 grid points equally
spaced in $\sqrt{r}$ 
          between $r_{\rm inner} = 10^7$ cm
and               $r_{\rm outer} = 1.5 \times  10^{11}$ cm.

The effect on the
evolution of the disk is quite pronounced.
Figures 2 and 3 show the time dependent accretion disk
evolution both with and without evaporation.
A purely viscous decay 
  with $\alpha_{\rm hot}=0.1$
  and disk parameters
appropriate for A0620-00
    gives a slower-than-exponential
decay, with a decay time constant  $\tau_e\approx 300$ d.
The solid line in Fig. 2 reveals the exponential character
of the decay if one has weak evaporation.
   In this model we reduce the amplitude of the 
evaporation  calculated from the CLOUDY models
   by a factor of 30.
       The time constant for the decay of ${\dot M}_{\rm inner}$
which powers  the soft X-ray flux
in this model is $\tau_e\approx 80-100$ d.
Fig. 3 shows the case for stronger evaporation
$-$ for which the amplitude of the evaporation  
                  calculated from the CLOUDY models
  is only reduced by a factor of 10 $-$ so that
$\tau_e\approx 20-30$  d.
   One now 
        sees a secondary maximum
  in the light curve of ${\dot M}_{\rm inner}$
        corresponding to the
 evacuation of the innermost disk, and its subsequent refilling.
The rapid rise  time for the secondary maximum is
a natural artifact of the fast viscous time at 
     $\sim10^8-10^9$ cm.
  The entire accretion disk is not able to 
re-adjust quasi-statically to the
persistent mass loss. 
    The viscous time scale,
which is equivalent to the time scale for the surface
density to adjust to strong perturbations,
is initially  $\tau_v \simeq 
      10^6$ s $(r/10^{10} \  {\rm cm})^{1.25}$
in
     our calculations, whereas the mass loss time
scale $\tau_{\dot M}$
     has about the same form but is roughly
$30$ times  slower (in our ``optimal'' model). 
 As evaporation proceeds near the inner edge,
  however,
              $\tau_v(r_{\rm inner})$ increases more rapidly than
       $\tau_{\dot M}(r_{\rm inner})$,  and at some point
   $\tau_{\dot M} < \tau_v$ 
  so that evaporation wins.

\section { DISCUSSION AND CONCLUSION  }
    
We have examined the decay from outburst
  in X-ray novae both with and without evaporation
of the inner disk. 
   In our model with $\alpha_{\rm hot} = 0.1$
(as inferred from dwarf novae), the decay time  $\tau_e$
varies from $\sim260$ d to $\sim340$ d over the 250 d
time span of our run. This time scale is 
about a factor 10 slower than observed, and
     would be expected in the scenario of King \& Ritter (1998)
in which the disk is held in the hot, ionized state
by strong irradiation expected in a soft X-ray transient.
Increasing $\alpha_{\rm hot}$ to 1 reduces the time
scale $\tau_e$
     associated with mass loss onto the central BH,
but also increases the deviation from exponentiality:
over the 250 d of the model, the $e-$folding time
scale associated with the decrease in ${\dot M}_{\rm inner}$
varies from $\sim40$ d to $\sim120$ d.
  Observations also exclude this model.
For models including evaporation during outburst,
   where the evaporation  is parametrized 
   using steady state computations
    of  the Shaviv \& Wehrse 
   instability,
%
     evaporation
is dominant over the viscous evolution
in the sense that the rate of removal
of material from the  inner disk
             leads to an $e-$folding
decay rate associated with the loss of material
from the inner accretion disk which is
$\ga 10$ times faster than that to due viscous 
evolution alone.
   More importantly,  the
decline maintains an exponential shape over
$\ga 2$ orders of magnitude
        in ${\dot M}_{\rm inner}$.

An unexpected consequence of the depletion
of matter from small radii in the disk
   is the effect of 
the evacuation
and subsequent refilling  of material
in the innermost   disk.
   This produces a  small increase
in ${\dot M}_{\rm inner}$, followed
by the   resumption of a rate of decrease
equal to what it had been earlier, 
   which seems to account 
  naturally for 
    the secondary  maxima.
The time constant associated with the post-secondary
maximum decay
  for the run shown in Fig.  3  
     is actually slightly  less
than that associated with the pre-secondary
maximum decay, as was observed in the 1975 outburst
of A0620-00 (see Fig. 1 of Kaluzienski et al. 1977).
   Also, 
   the fast  rise time of $\sim1-3$ d for the
 secondary maximum in our model
        represents
the viscous evolution time at
small radii  $\sim10^8$ cm
in the disk. By contrast, 
    the previous models of the secondary 
maxima  by Chen et al. (1993) 
and Augusteijn et al. (1993) which  invoke
the addition of extra material from the secondary
would give very slow secondary  maxima 
   with rise times of  $\sim1$ yr,
because the matter added to the outer
disk  at $\sim 10^{11} $ cm
would require a long time to enhance the accretion
   rate at
   small radii where X-rays are produced.
 Another point of consistency with observations
  in our model for the secondary maximum is that,
insofar as its root cause 
is a readjustment of the surface density
distribution in the inner disk,
the impact on the $V$ light curve is minimal.
  The $V$ band flux is heavily weighted by
contributions from the outer disk radii (Cannizzo 1996),
and therefore it  responds on a much longer time scale
to processes occurring at the inner edge $-$
effectively smearing out their effects.



We thank Clara Hughes in the
Laboratory for Astronomy and Solar Physics
and Tom Cline in the 
Laboratory for High Energy Astrophysics (LHEA)
at Goddard Space   Flight Center (GSFC)
for generous use of their DEC alpha
workstations.
We acknowledge useful discussions
  with
 M. de Kool,
  G. Ferland,
   I. Hubeny,
  T. Kallman,
  J.-P. Lasota,
    S. Nayakshin,
        E. Vishniac,
   and J. C. Wheeler.
   JKC
 was supported 
through the long-term 
scientist program under
 the Universities Space Research Association
(USRA contract NAS5-32484) 
 in the LHEA at GSFC.

\vfil\eject

\vfil\eject
\centerline{ FIGURE CAPTIONS }

Figure 1. The vertical structure of the optically thin part
of the accretion disk overlying the photosphere.
The input parameters are $M_1=10\msun$, $r=10^{9.5}$ cm,
and $\alpha_{\rm chromosphere} = 0.3$.
Five  curves are shown, for  ${\dot M} = 10^{14.5}$ g s$^{-1}$
through  ${\dot M} = 10^{18.5}$ g s$^{-1}$
 in steps   of 1.0 dex. 
  Shown are the logarithms of 
             gas pressure $P_{\rm gas}$({\it top panel}),
             density $\rho$ ({\it middle panel}),
  and               $2\pi r^2\rho c_s$  ({\it bottom panel})
which has units of g s$^{-1}$  and provides
  a measure of the expected outflow
  in regions where hydrostatic equilibrium
cannot be maintained (Czerny \& King 1989).


Figure 2. The time dependent evolution of the
  accretion disk both with and without
   evaporation.
  Shown are the absolute $V$ mag for a face-on disk
({\it top panel}), the mass of the disk in units of
$10^{25}$ g ({\it second panel}),
     the logarithm of the rate of mass removal 
from the inner disk edge in  g s$^{-1}$ ({\it third panel}), 
and the locally defined $e-$folding decay time $\tau_e$
for $(dm/dt)_i$ ({\it bottom panel}).
  The three curves shown represent runs for
(i) a viscous decay with  $\alpha_{\rm hot}=0.1$
({\it dotted curve}), 
(ii) a viscous decay with  $\alpha_{\rm hot}=1$,
({\it dashed curve}), and
(iii) a viscous plus (weak) evaporative
     decay with  $\alpha_{\rm hot}=0.1$ ({\it solid curve}),
  where the strength of the evaporation
     has been reduced by 30 from what was computed using CLOUDY and
shown in Fig. 1.
 For these runs   no cooling front is present because
 of the strong irradiation.
  For the  viscous decay run  with $\alpha_{\rm hot}=0.1$,
 the associated decay time constant $\tau_e$  varies
from  $\sim260$ d to $\sim340$ d, whereas in the
run for $\alpha_{\rm hot}=1$,   $\tau_e$ increases
from $\sim40$ d to $\sim120$ d over the  period of 
evolution shown.  In the model 
           with weak evaporation
a closely  exponential
decay with $\tau_e\simeq80-100$ d results.

Figure 3. The evolution of the accretion 
   disk with stronger evaporation
   than what was shown in Fig. 2.
   The four panels are the same as in Fig. 2.
 For this run we reduce the  strength of the   evaporation
      by a factor of 10 from what was computed using CLOUDY
and shown in Fig. 1, resulting in a decay with 
                $\tau_e \simeq 20-30$ d. 
   By $t=100$ d the inner disk has evaporated, and material
  from   further out in the disk flows inward to fill the
cavity. 
 The post-secondary maximum decay
is exponential also, with
a  slightly faster time constant  than before.

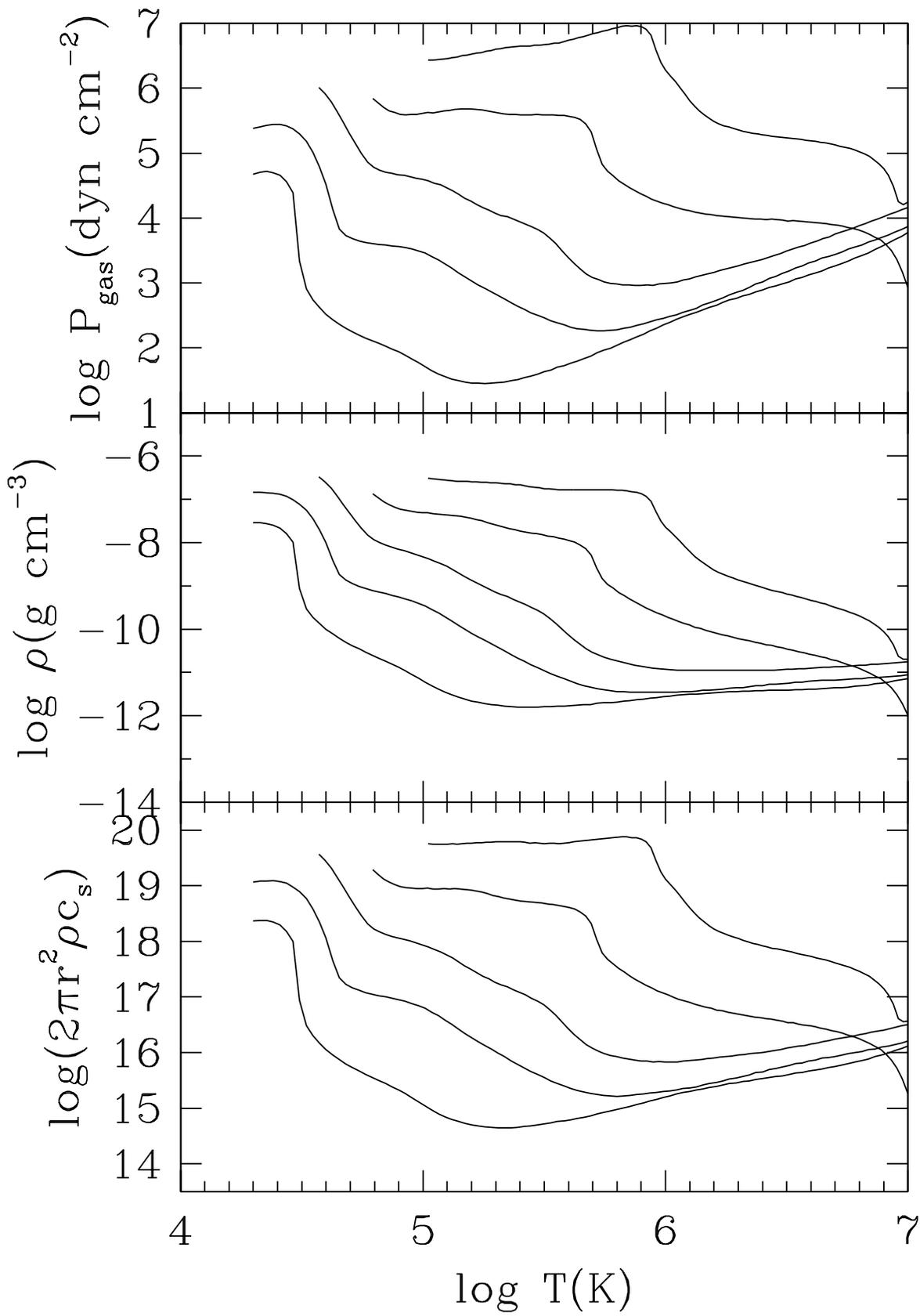
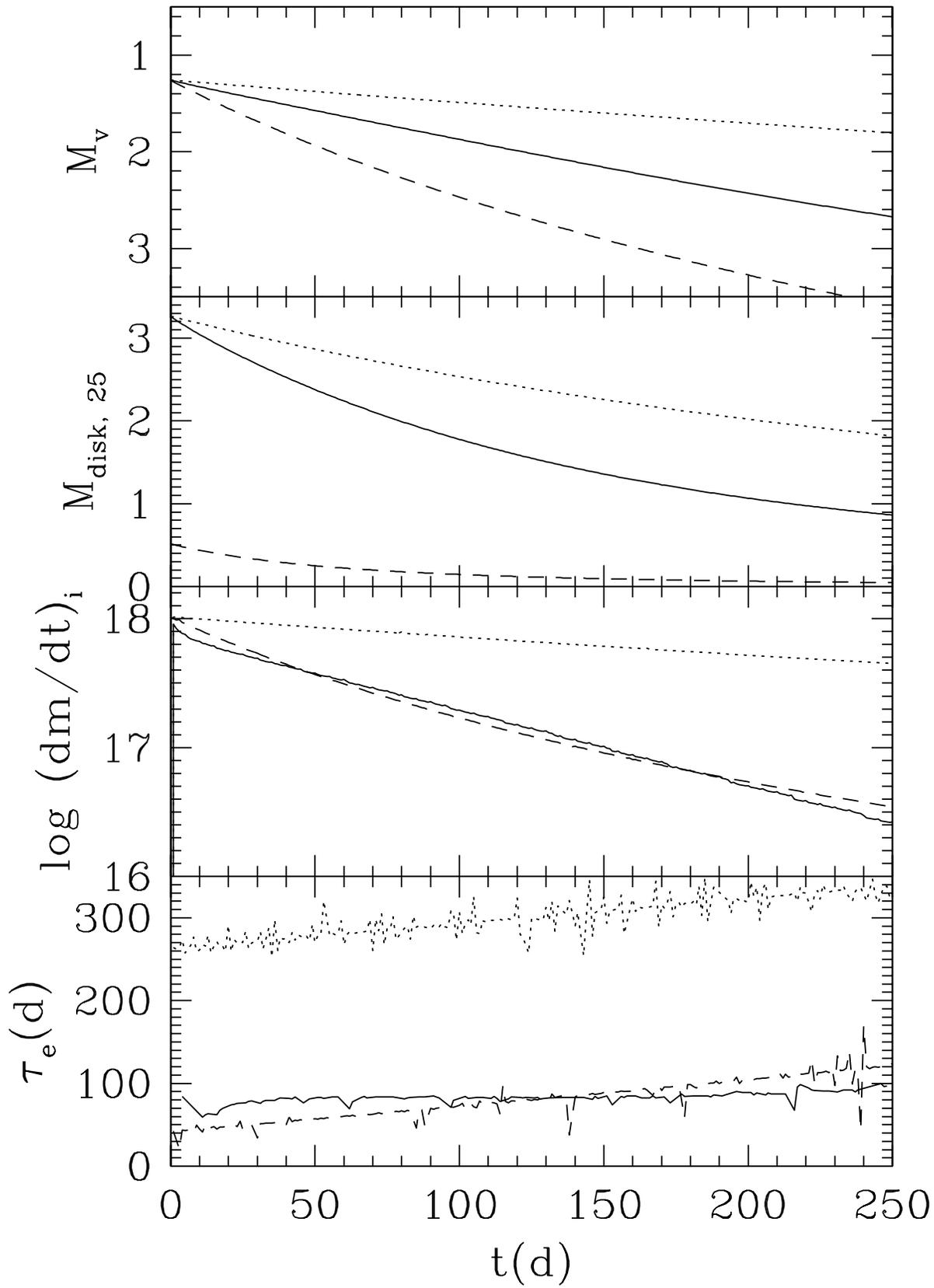
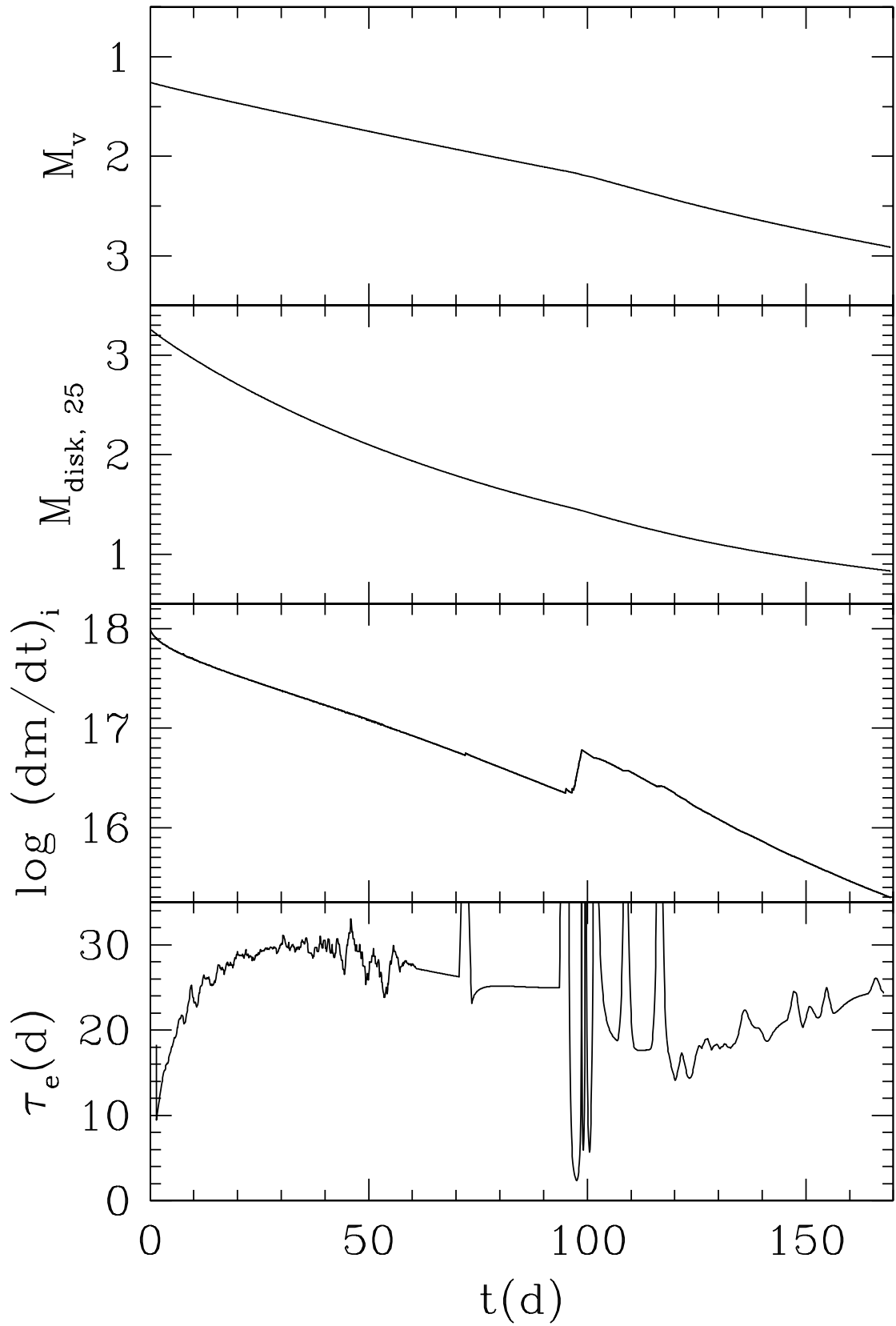
\end{document}